\documentclass[twocolumn]{aastex63}
\usepackage{lineno}
\usepackage{epstopdf}
\usepackage{amsmath}
\shorttitle{Picking-up Local GRBs}
\shortauthors{Wang et al.}

\begin{document}

\title{Picking-up Local GRB Candidates Based on Their Host Galaxies}

\correspondingauthor{J. Wang, Y. Xu}
\email{wj@nao.cas.cn, yxu@nao.cas.cn}

\author{J. Wang}
\affiliation{Key Laboratory of Space Astronomy and Technology, National Astronomical Observatories, Chinese Academy of Sciences, Beijing 100101,
People's Republic of China}
\affiliation{Guangxi Key Laboratory for Relativistic Astrophysics, School of Physical Science and Technology, Guangxi University, Nanning 530004, People's Republic of China}
\affiliation{GXU-NAOC Center for Astrophysics and Space Sciences, Nanning, 530004, People's Republic of China}

\author{Y. Xu}
\affiliation{Key Laboratory of Space Astronomy and Technology, National Astronomical Observatories, Chinese Academy of Sciences, Beijing 100101,
People's Republic of China}

\author{L. J. Chen}
\affiliation{Guangxi Key Laboratory for Relativistic Astrophysics, School of Physical Science and Technology, Guangxi University, Nanning 530004, People's Republic of China}

\author{C. Wu}
\affiliation{Key Laboratory of Space Astronomy and Technology, National Astronomical Observatories, Chinese Academy of Sciences, Beijing 100101,
People's Republic of China}
\affiliation{School of Astronomy and Space Science, University of Chinese Academy of Sciences, Beijing, People's Republic of China}

\author{L. P. Xin}
\affiliation{Key Laboratory of Space Astronomy and Technology, National Astronomical Observatories, Chinese Academy of Sciences, Beijing 100101,
People's Republic of China}

\author{E. W. Liang}
\affiliation{Guangxi Key Laboratory for Relativistic Astrophysics, School of Physical Science and Technology, Guangxi University, Nanning 530004,
People's Republic of China}
\affiliation{GXU-NAOC Center for Astrophysics and Space Sciences, Nanning, 530004, People's Republic of China}

\author{J. Y. Wei}
\affiliation{Key Laboratory of Space Astronomy and Technology, National Astronomical Observatories, Chinese Academy of Sciences, Beijing 100101,
People's Republic of China}
\affiliation{School of Astronomy and Space Science, University of Chinese Academy of Sciences, Beijing, People's Republic of China}






\begin{abstract}

Rapid identification of candidates of high-value gamma-ray bursts (GRBs), including both high-$z$ and
local events, is crucial for outlining subsequent observational strategy.
In this paper, we present a model that enables an on-duty astronomer to rapidly
identify candidates of local GRBs prior to spectroscopy, 
provided that these events have been localized at an arcseconds precision.
After taking into account the mass distribution of the host galaxies of GRBs, the model 
calculates the two-dimensional cross-match probabilities between 
a localized GRB and its surrounding nearby galaxies, and then returns the best match with the 
highest probability. The model is evaluated not only by the observed GRB sample with redshifts up to $z=4$,
but also through the simulated GRB samples. 
By using the recently published GLADE+ galaxies catalog with a completeness of 95\% up to 500Mpc, along with 
the NED-LVS catalog, the 
Precision and Recall of the model are determined to be 0.23-0.33 and 0.75, respectively, at the best performance.
A dedicated web service, which will be integrated into the SVOM Science User Support System,
has been developed to deploy the model.

\end{abstract}



\section{Introduction} \label{sec:intro}

Gamma-ray bursts (GRBs) are the most powerful explosions occurring in the universe (e.g.,
Hjorth \& Bloom 2012). There is compelling observational evidence supporting that
the GRBs originate from either the core-collapse of young massive stars ($\geq25M_\odot$;
e.g., Woosley \& Bloom 2006 and references therein) or the merger of neutron star binaries
(e.g., Berger 2014 and references therein).

Due to these widely accepted origins, GRBs are believed to occur from local to distant universe up to a
redshift of $\sim$20 (e.g., Lamb \& Reichart 2000). Although high-$z$ GRBs are fascinating in the
exploration of the early universe (e.g., Tagliaferri et al. 2005; Greiner et al.
2009; Tanvir et al. 2009, 2018; Melandri et al. 2015; Cucchiara et al. 2011), the local, and usually bright,
GRBs are
still important because they allow us to explore the nature of the progenitor and environment of a
GRB comprehensively through either high-quality afterglow spectra or well resolved host galaxy.

In fact, the aforementioned collapse and merger scenarios are confirmed by
the detection of a supernova (SN) or a kilonova associated with a GRB, respectively.
On the one hand, mainly because of the faintness of the associated SNe, there are, so far, only  30 spectorscopically confirmed GRB-SNe (e.g., Cano et al. 2017a, b;
Wang et al. 2018; Melandri et al. 2019;
Hu et al. 2021; Blanchard et al. 2023; Srinivasaragavan et al. 2024; Gompertz et al. 2024),
in which more than one third of the cases have $z<0.2$.
The core-collapse scenario is indirectly supported by the fact that GRBs with $z<1.2$
are found to be concentrated on the very bright regions of their host galaxies (e.g. Fruchter et
al. 2006). On the other hand, the kilonova produced by a merger of a
neutron star binary has been confirmed by spectroscopy in only two nearby GRBs,
GRB~170817A and GRB~230307A, within a distance of 300Mpc (e.g., Pian et al. 2017; Shappee et al. 2017;
Levan et al. 2024). In addition,
multi-epoch, high-quality afterglow spectra of bright (usually not very distant) GRBs
have been used to
distinguish different scenarios of the interaction between the GRB's relativistic jet and
its environment, and to
determine the distance between the GRB and the absorbing material by the variability of the
fine-structure lines (e.g., Vreeswijk et al. 2007, 2013; D'Elia et al. 2010, 2014; Kruhler et al. 2013;
Hartoog et al. 2013; Wiseman et al. 2017; Pugliese et al. 2024).

Because GRBs usually decay rapidly at their early time, a fast assessment 
of their distance in prior spectroscopy is quite crucial for mapping out the subsequent 
observational strategy.
This is, however, a hard task for local GRBs. Not as the high-$z$ cases, the afterglow spectral-energy-distribution
is useless in the redshift assessment for low-$z$ GRBs because of the lack of the
Lyman $\alpha$ break feature (e.g., Wang et al. 2020).

Due to the rapid progress in the time-domain astronomy, 
a few studies have been recently carried out by aiming at 
the identification of the host galaxy candidates of extragalactic transients, 
such as GRBs, fast radio bursts and gravitational wave events, by an approach of 
probability (e.g., Aggarwal et al. 2021; Ducoin 2023; Demasi et al. 2024). 
In this paper, by involving the mass of GRB's host galaxy,
we propose a new model that enables us to select local GRB candidates ($z<0.1$) by
a probability that is assessed by a two-dimensional
cross-match between GRB afterglows and nearby galaxies with
either spectroscopic or photometric redshifts.

The paper is organized as follows. The proposed whole schema of picking-up local
GRBs in the SVOM mission\footnote{
SVOM, launched in 2024, June 22, is a Chinese-French space mission dedicated to the detection and study of GRBs.
We refer the readers to Atteia et al. (2022) and the white paper given by Wei et al.
(2016) for the details.} are
presented in Section 2. Section 3 describes the conception of the proposed model in details.
An assessment of the model, along with the used GRB and nearby galaxy samples, are
presented in Section 4. Section 5 gives the conclusion and application in the SVOM mission. A $\Lambda$CDM cosmological model with parameters
$H_0 = 67.4\ \mathrm{km\ s^{-1}\ Mpc^{-1}}$, $\Omega_m= 0.315$, and $\Omega_\Lambda= 0.685$
(Planck Collaboration et al. 2016)
is adopted throughout the paper.

\section{Schema of Picking-up Local GRB Candidates in SVOM} \label{sec:style}

Figure 1 shows the flow chart of picking-up local GRB candidates
based on the observations taken by the instruments of the SVOM mission. 
As shown in the figure, only based on the trigger and follow-up information down-loaded through the VHF network, two independent approaches are designed to select local GRB candidates. 
One approach is based on 
the high-energy prompt emission of a GRB. After excluding the possibility of a high$-z$ GRB
candidate ($z>4$), the other approach is based on a 
two-dimensional cross-match between a GRB and surrounding nearby galaxies,
provided that the GRB has been localized at arcseconds level.

\rm
\begin{figure*}
   \centering
   \plotone{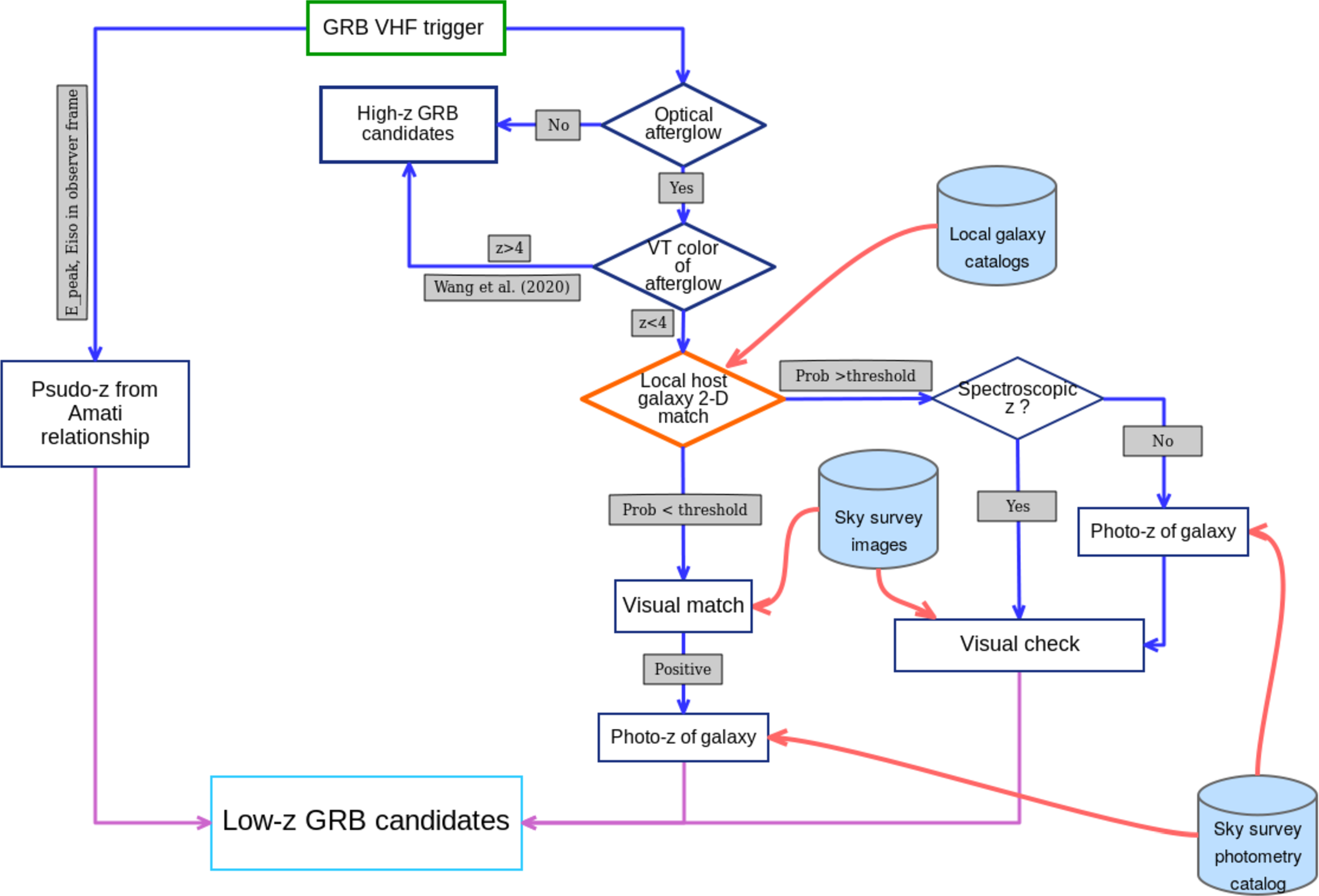}
   \caption{Flow chart of picking-up local GRB candidates in the SVOM mission only by the 
   VHF trigger and follow-ups prior to spectroscopy. The
   left and right flows depend on the GRB's prompt emission and the localization of afterglows
   at arcseconds level, respectively.
   The data flows and final results are denoted by the blue and magenta arrows. Three external
   sources, including galaxy catalogs and deep image surveys, are needed to realize the
   flows. The process related with this paper is emphasized by orange.
}
\label{Fig1}
\end{figure*}

\section{The Model}

By involving the mass of a galaxy,
the probability $P_{\mathrm{gal}}$ that
a GRB, 
with a localization at arcseconds level and without a redshift measurement,
resides in the galaxy (with $z_{\mathrm{gal}}$) is calculated as

\begin{equation}
   P_{\mathrm{gal}}= P_{\mathrm{M}}(M_\star/M_\odot)\times P_{\mathrm{A}}(r_{\mathrm{A}})  
\end{equation}
where $M_\star$ is the total stellar mass of the galaxy and $r_{\mathrm{A}}$ the linear
distance between the GRB and the center of the galaxy if the GRB has $z_{\mathrm{gal}}$.
The probability density functions 
$P_{\mathrm{M}}(M_\star)$ and $P_{\mathrm{A}}(r_{\mathrm{A}})$ are
normalized and described as follows.

There are plenty of studies suggesting that GRB's host galaxies are biased towards less massive and
metallicity poor end (e.g., Vergani et al. 2015; Perley et al. 2016a,b). We determine the
$P_{\mathrm{M}}(M_\star/M_\odot)$ from the sample reported in Perley et al. (2016b). 
With the \it Swift \rm GRB Host Galaxy Legacy Survey, 
the sample is composed of the host galaxies of 119 bursts.
The stellar masses of these galaxies are converted from the corresponding rest-frame 
near-infrared (NIR) luminosities measured in
the deep images taken by \it Spitzer Space Telescope. \rm The distribution of the stellar masses
of these GRB host galaxies
is shown in the left panel of Figure 2. Modeling the distribution by a Gaussian function

\begin{equation}
  P_{\mathrm{M}}(M_\star/M_\odot)= \frac{1}{\sqrt{2\pi}\sigma} e^{-\frac{[\log(M_\star/M_\odot)-\mu]^2}{2\sigma^2}}
\end{equation}
yields $\mu = 9.73\pm0.06$ and $\sigma = 0.64\pm0.06$ (the solid blue line overplotted in Figure 2).

We calculate $P_{\mathrm{A}}(r_{\mathrm{A}})$ by a piecewise exponential distribution

\begin{equation}
  P_{\mathrm{A}}(r_{\mathrm{A}}) = \begin{cases}
                                    1, & if \ r_{\mathrm{A}} \leq r_{\mathrm{e}}\\
                                    \frac{1}{r_{\mathrm{e}}}e^{-(r_{\mathrm{A}}/r_{\mathrm{e})}} , & if \ r_{\mathrm{A}} > r_{\mathrm{e}} 
                                   \end{cases}
\end{equation}
after converting the angular distance $\theta_{\mathrm{A}}$ to $r_{\mathrm{A}}$ by $r_{\mathrm{A}}=\theta_{\mathrm{A}} d_{\mathrm{A}}$.
$d_{\mathrm{A}}=d_{\mathrm{L}}/(1+z_{\mathrm{gal}})^2$ is the angular diameter distance, and $d_{\mathrm{L}}$ the
luminosity distance.
The parameter $r_e$ is the physical size of a galaxy that can be estimated from the stellar mass according to
the observed $R-M$ relation given in Shen et al. (2003). Based on 140~000 SDSS galaxies,
the authors reported a dependence of $R_{50}$, the radius enclosing 50\% of the Petrosian
flux\footnote{According to the definition given in Petrosian (1976), the Petrosian flux
is $\sim$98\% and $\sim80$\% of the total flux for an exponential and a de Vaucouleurs profiles, respectively.}, on
luminosity, stellar mass and morphological type. Specifically, the early type galaxies
follow:

\begin{equation}
  R^{\mathrm{early}}_{50} = 3.47\times10^{-5}\bigg(\frac{M_\star}{M_\odot}\bigg)^{0.56}\ \mathrm{kpc}
\end{equation}
and the late type galaxies have

\begin{equation}
  R^{\mathrm{late}}_{50} = 0.1\bigg(\frac{M_\star}{M_\odot}\bigg)^{0.14}\bigg(1+\frac{M_\star}{M_0}\bigg)^{0.39}\ \mathrm{kpc}
\end{equation}
where $M_0=3.98\times10^{10}M_\odot$.
Because the morphology is not always available in a galaxy catalog,
we adopt $r_{\mathrm{e}}=\max(R^{\mathrm{early}}_{90},R^{\mathrm{late}}_{90}$) in
the current study to avoid an underestimation of the galaxy size, where
$R_{90}$ is the radius enclosing 90\% of the Petrosian flux.
The ratio $R_{90}/R_{50}$, i.e., the concentration index, is typically $\sim2.3$
and $\sim3.3$ for an exponential and a de Vaucouleurs profiles, respectively.

\begin{figure*}
   \centering
   \begin{minipage}{0.49\linewidth}
      \centering
      \includegraphics[width=1.0\linewidth]{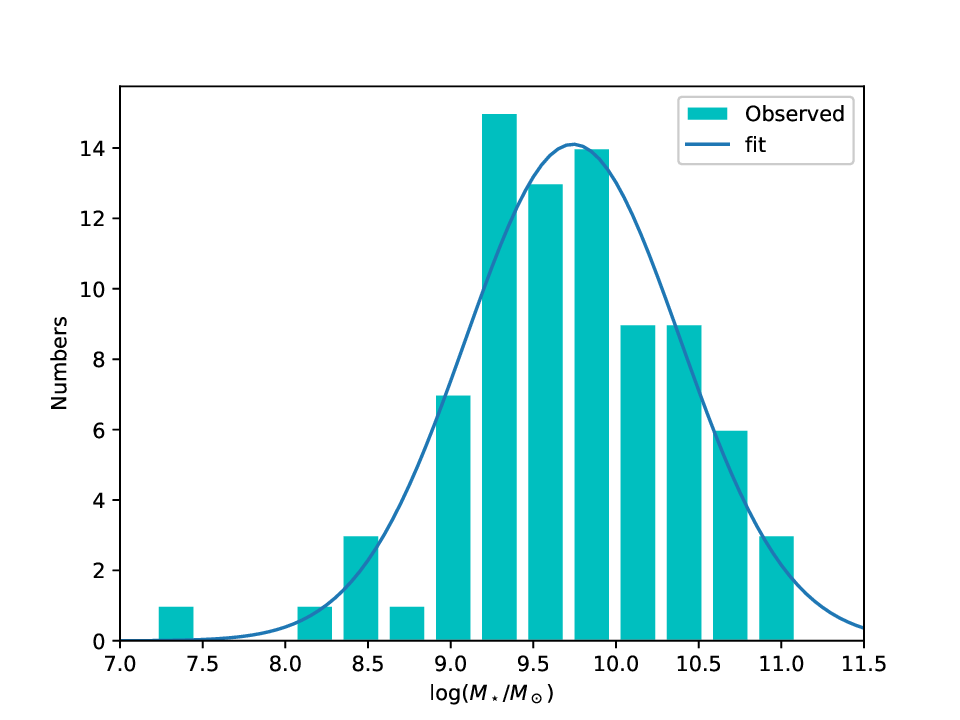}
   \end{minipage}
   \begin{minipage}{0.49\linewidth}
      \centering
      \includegraphics[width=1.0\linewidth]{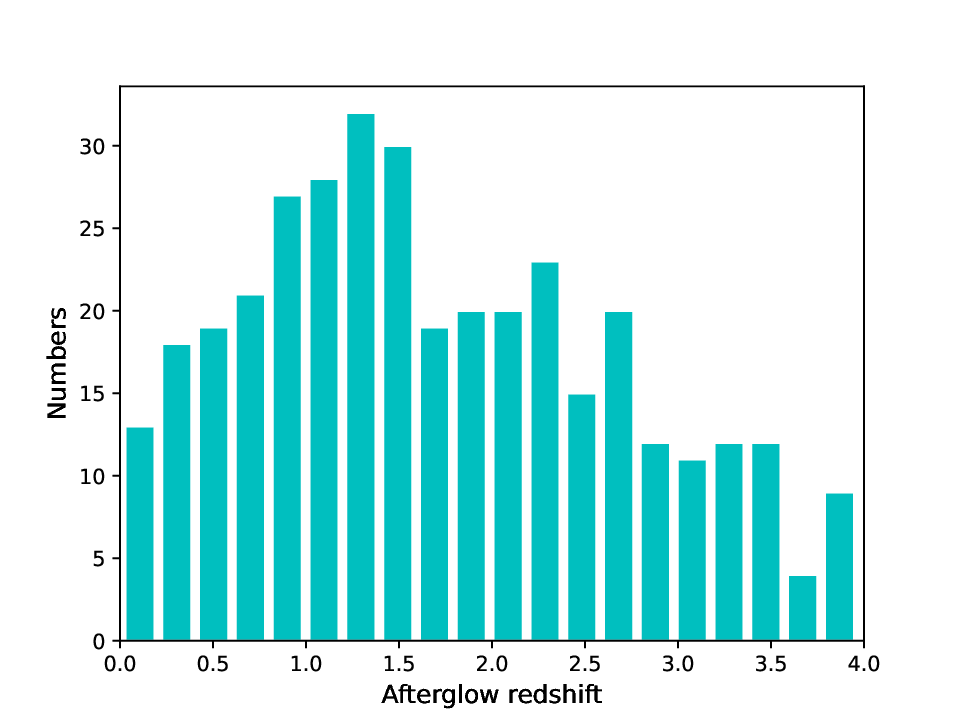}
   \end{minipage}
   \caption{\it Left panel: \rm Distribution of the stellar mass of the host galaxies of a sample of
   GRBs extracted from Perley et al. (2016b). The best fit Gaussian function with $\mu=9.73\pm0.06$ and
   $\sigma=0.64\pm0.06$ is overplotted by the
   blue solid line. \it Right panel: \rm Distribution of the spectroscopic redshifts  of a
   sample of GRBs ($z\leq4.0$) complied from the literature.
}
\label{Fig2}
\end{figure*}

For a given GRB localized at arcseconds level, each nearby galaxy surrounding the GRB
is assigned an association probability calculated by following  Eq. (1). 
The galaxy with the highest association probability $P_{\mathrm{max}}$ is then extracted. 
If $P_{\mathrm{max}} > P_0$, where $P_0$ is the probability threshold that is determined by the model 
evaluation and tuning described in the next section, we accept that the galaxy is 
the most probable host galaxy of the GRB. 
One should bear in mind that the ``acception''
does not mean the galaxy with $P_{\mathrm{max}} > P_0$ is just the true host, simply because
the associated probability is calculated in two-dimension. 
Due to the absence of radial distance of GRBs, there is likely considerable contamination caused by
accident matches between the local galaxies and distant GRBs. 

\section{Model Evaluation}

In this section, our goal is to determine the optimal performance of the proposed model,
which is achieved by tuning the probability threshold $P_0$ based on known nearby galaxy catalogs. 
Both observed and simulated GRB samples are used in the evaluation.

\subsection{The Nearby Galaxy Catalogs}

Two nearby galaxy catalogs are considered in our model evaluation. 
One is the GLADE+ galaxy catalog\footnote{http://glade.elte.hu/.} recently published by Dalya et al. (2022). 
The catalog is developed for searching for electromagnetic
counterparts of advanced gravitational detectors by combining six independent catalogs:
the GWGC, 2MPZ, 2MASS XSC, HyerLEAD, WISExSCOPZ and SDSS-DR16 quasar catalogs.
It contains $\sim22.5$ million galaxies and $\sim$750~000 quasars, achieving full completeness
in the $B-$band within a luminosity distance of 47Mpc
(corresponding to $z\approx0.01$). A completeness of $\sim90$\% can be
achieved at a distance of $\sim500$Mpc ($z\approx0.1$) when the NIR $W1-$band is involved.
In addition, the catalog provides the total stellar mass estimated from the $W1-$band brightness.

The other is the NED Local Volume Sample (NED-LVS)\footnote{https://ned.ipac.caltech.edu/NED::LVS/.} that contains $\sim2$ million galaxies with 
a distance up to 1~Gpc (Cook et al. 2023). Based on the near-IR luminosities,
the catalog has a completeness of $\sim100\%$ at a distance of 30Mpc. The 
completeness decreases to $\sim70\%$ up to 300Mpc, and remains $\sim100\%$ out 
to $\sim400$Mpc for bright galaxies ($\geq L_\star$, $L_\star$ is the
characteristic luminosity (or ``knee'') of the luminosity function).


\subsection{Model Evaluation}

With the two nearby galaxy catalogs described above, the model presented in Section 3 is
evaluated by both observed and simulated GRB samples.

\subsubsection{Evaluation from the Observed GRB Sample}

 
A sample of GRBs with spectroscopic redshifts is compiled from literature.
An upper limit of $z=4.0$ is adopted for the sample for two reasons. On the one hand,
compared to high-$z$ objects, the Lyman-$\alpha$ break feature is typically weak for objects
with $z<4$ due to the deficient Lyman-$\alpha$ optical depth. On the other hand, there is
proven technique for identifying high-$z$ GRB candidates by multi-bands photometry
(e.g., Wang et al. 2020). In total, the sample is composed of 365 GRBs with
spectroscopic redshifts $z<4.0$.
The right panel of Figure 2 shows the distribution of the redshifts of
the GRB sample used in the current study.

With the GRB sample and the nearby galaxy catalogs,
we at first calculate the predictions of our model by following the method
described in Section 2. For each GRB, the calculation is performed for the
galaxies within a circle of 10\arcmin\footnote{This large circle is adopted to avoid missing of very
nearby and bright galaxies in the calculation. Except that, there is no impact on the 
final model performance due to the rapid decay of the exponential distribution in Eq (3).}.
In the GRB sample, there are
343 bursts that have least one galaxy within a circle of 10\arcmin.
In the GLADE+ catalog, there are a few galaxies without estimated stellar mass.
A default maximum probability of $P_{\mathrm{M}}=0.17$ is assigned to these galaxies. 
In both GLADE+ and NED-LVS catalogs, the galaxies
without either spectroscopic or photometric redshifts are excluded in the model calculation,
after all the fraction of these objects is quite small. The value of $P_{\mathrm{max}}$ 
is then determined for each GRB by following the method described at the end of Section 3.

In order to extract ``true'' matches between the 343 GRBs and local galaxies,
a revised model in three-dimensional space is additionally  
calculated by including the redshifts or distances of the objects listed in the GRB sample.
Specifically, for a given GRB, the probability $P'_{\mathrm{gal}}$ that a nearby galaxy is associated with it in three-dimensional space can be calculated as 
\begin{equation}
  P'_{\mathrm{gal}}= P_{\mathrm{M}}(M_\star)\times P_{\mathrm{A}}(r_{\mathrm{A}})\times P_{z}(\Delta z) 
\end{equation}
where $P_{\mathrm{M}}(M_\star)$ and $P_{\mathrm{A}}(r_{\mathrm{A}})$ are given in Eqs. (2) and (3), respectively.
Similar as $P_{\mathrm{A}}(r_{\mathrm{A}})$, a piecewise exponential distribution

\begin{equation}
  P_{z}(\Delta z) =\begin{cases}
                      1, & if \ \Delta z \leq \Delta z_0\\
                      \frac{1}{\Delta z_0}e^{-\Delta z/\Delta z_0} , & if \ \Delta z > \Delta z_0
                                   \end{cases}
\end{equation}
is adopted to calculate the probability in three-dimensional space, where
$\Delta z = |z_{\mathrm{GRB}}-z_{\mathrm{gal}}|$ is the redshift difference between the GRB and
galaxy. $\Delta z_0$ is fixed to be 0.005 in the calculation, which corresponds to a velocity
difference of 1,500$\mathrm{km\ s^{-1}}$. This velocity difference is large enough to
account for the escaping
velocity of a few$\times10^2\mathrm{km\ s^{-1}}$ of a galaxy. For example, the escaping
velocity of the Milk Way is between 492 and 594$\mathrm{km\ s^{-1}}$.

After calculating the association probability in three-dimensional space, 
the galaxy with the highest association probability, denoted by $P'_{\mathrm{max}}$, is then
identified for each GRB listed in the sample. \rm
The probability $P'_{\mathrm{max}}$ calculated in three-dimensional space is highly useful for identifying 
genuine matches between the GRBs and their host galaxies.
When the GLADE+ catalog is adopted, 
the distribution of $P'_{\mathrm{max}}$ is in fact found to be
concentrated at the extremely small value end. The median and geometric mean
of $P'_{\mathrm{max}}$ are close to zero \rm for the 343 GRBs. Among the 343 GRBs,
there are only nine ones with $P'_{\mathrm{max}}>1\times10^{-4}$.
By visually examining the sky images one by one, the truth of the cross-match is
validated for 7 out of the 9 GRBs.
The details of the validated cross-match of the 7 GRBs is tabulated in Table 1. Figure 3 shows the
corresponding sky images.

\begin{deluxetable*}{ccccccccc}
        \centering
        \tablecaption{The validated seven cross-match between the GRBs and galaxies listed in the GLADE+ catalog.}
        \footnotesize
        \label{tab:example_table}
        \tablehead{
        \colhead{GRB} & \colhead{Host galaxy} & \colhead{GLADE+ No.} & \colhead{$B_\mathrm{abs}$} & \colhead{$d_{\mathrm{L}}$} & \colhead{$M_\star$} & \colhead{$r_{\mathrm{A}}$} & \colhead{$\Delta z$} & \colhead{$P_{\mathrm{max}}$}\\
          &  & \ & \colhead{mag} & \colhead{Mpc} & \colhead{$10^{10} M_\odot$} & \colhead{\arcsec} &  & \\
        \colhead{(1)} & \colhead{(2)} & \colhead{(3)} & \colhead{(4)} & \colhead{(5)} & \colhead{(6)} & \colhead{(7)} & \colhead{(8)} & \colhead{(9)}
        }
        \startdata
                  060505 & \dotfill & 614911 &  -19.7 & 415 & 0.5 & 4.8 & 0.0012 & 0.1696  \\
                  171205A & 2MASX~J11093966-1235116 & 1198372 & -20.8 & 175 & 1.0 & 6.5 & 0.0002 & 0.1570 \\
                  170817A & NGC~4993 & 1151336 & -19.9 & 41 & 2.2 & 10.0  & 0.00005 & 0.1105\\
                  190829A & SDSS~J025810.28-085719.2 & 1246359 & -21.3 & 371 & 4.0 & 11.6 & 0.0005 & 0.0705\\
                  980425 & ESO184-G82 & 1265105 & -19.3 & 67 & 0.1 & 11.9 & 0.0063 & 0.0239\\
                  191019A & \dotfill & 20916216 & -19.7 & 1206 & 4.0 & 0.1 & 0.0142 & 0.0047\\
                  130702A & \dotfill & 14619972 & -20.2 & 843 & 6.0 & 7.0 & 0.0245 & 0.0004\\
        \enddata
        \tablecomments{Column (1): GRB name; Column (2): Host galaxy name, if possible; Column (3):
        The ID of the galaxy in the GLADE+ catalog; Column (4): Absolute magnitude in $B-$band of
        the galaxy; Column (5): Distance of the galaxy in unit of Mpc; Column (6): Stellar mass
        of the galaxy; Column (7): Angular offset (in unit of arcsec) between the GRB's optical afterglow and the center of the galaxy;
        Column (8): Redshift difference between the GRB and galaxy; Column (9): The best match
        probability $P_{\mathrm{max}}$.}
\end{deluxetable*}

\begin{figure*}
   \centering
   \plotone{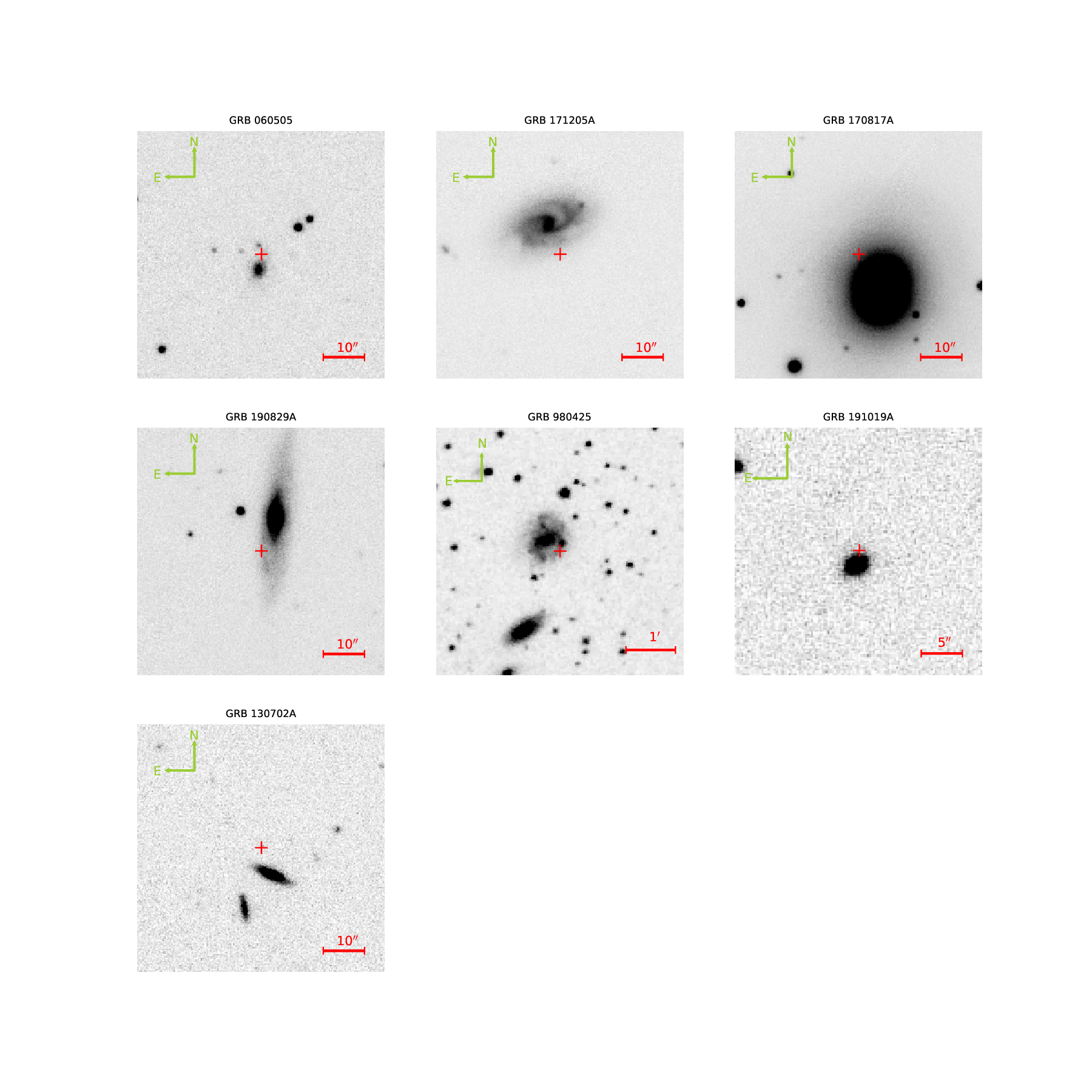}
   \caption{Sky images of the validated seven cross-matches between GRBs and galaxies. All the
   images obtained in the $i-$band are extracted from the Pan-STARR Date Release 1
   (Flewelling et al. 2020), except for GRB~980425, whose $R-$band image is extracted from the DSS2 survey. The size of each image is 1\arcmin$\times$1\arcmin, except for the cases of GRB~980425 and GRB~191019A whose images have
   a size of 5\arcmin$\times$5\arcmin\ and 30\arcsec$\times$30\arcsec\, respectively. In each panel, the position of the GRB's afterglow is
   at the center and marked by a red cross. The north and east are at top and left, respectively. 
}
\label{Fig1}
\end{figure*}

 
After identifying of the 7 genuine cross-matches, 
the performance of the model is assessed by calculating the
confusion matrix at various probability threshold $P_0$. The matrix is composed of four elements: i.e., the rates of
\it true \rm Positive (TP), \it false \rm Positive (FP),  \it false \rm Negative (FN) and
\it true \rm Negative (TN). Based on the matrix, 
panels (a) and (b) in Figure 4 shows the dependence of the various assessments, 
including the
Accuracy, Precision, Recall and F1-score\footnote{The parameter of accuracy measures the performance of a model, and is defined as the ratio of total correct instances
to the total instances
\begin{equation}
  \mathrm{ACC = \frac{TP+TN}{TP+FP+FN+TN}}
\end{equation}
The parameter of precision measures accuracy of \it true \rm positive predictions, and is defined as the
ratio of \it true \rm positive to the total number of positive predictions
\begin{equation}
  \mathrm{Precision = \frac{TP}{TP+FP}}
\end{equation}
The parameter of recall measures the effectiveness of a model in identifying all relevant instances from dataset,
and is defined as
\begin{equation}
  \mathrm{Recall = \frac{TP}{TP+FN}}
\end{equation}
The F1-score evaluates the overall performance of a model assessed by the harmonic mean of precision and
recall
\begin{equation}
  \mathrm{F1 = \frac{2\cdot Precision\cdot Recall}{Precision+Recall}}
\end{equation}
The value of F1-score ranges from 0 to 1. It reaches $F1=1$ when both Precision and Recall are unity.
However, $F1=0$ if one of the two parameters is zero. The factor of 2 in the numerator ensures
the calculated F1 value are between 0 and 1.},
on the probability threshold $P_0$, when the GLADE+ and NED-LVS
catalogs are used, respectively.
The assessments are calculated by the
python/\it sklearn \rm package (Pedregosa et al. 2011).

One can see from both panels that although both Accuracy and Precision increase with the threshold $P_0$, 
there is a consequent decline in the Recall. It means that at a low threshold $P_0$ 
despite the fact that local GRBs can be almost completely predicted by the model, there is still significant contamination from accidental cross-matching between distant GRBs and local galaxies. 
On the contrary, when a high threshold $P_0$ is adopted, although the contamination is greatly
reduced, a substantial number of local GRBs are overlooked in the prediction.

The F1-score is a good assessment providing
a balance between Accuracy and Recall, in which the optimal performance of a model can be determined at
the peak of F1-score. The determined optimal performance of our model is tabulated in Table 2.
Based on the values listed in the table, the model has the optimal performance at  
the probability threshold $P_0=0.05-0.09$. At the optimal performance, there are
23-33 true positives in 100 local GRB candidates predicted by the model. \rm At the same time, 
4-11 true local GRBs are, however, missed by the model as estimated from the 
values of Recall.

\begin{table*}
        \centering
        \caption{The best performance of the proposed model.}
        \label{tab:example_table}
        \begin{tabular}{cccccc} 
                \hline
                \hline
                  $P_0$ & Accuracy & Precision & Recall & F-measurement  & Galaxy catalog \\
                  (1) & (2) & (3) & (4) & (5) &  (6)\\
                  \hline
                  \multicolumn{6}{c}{Observed GRBs}\\
                  \hline
                  $0.060-0.070$ & 0.94 & 0.23  & 0.75 &  0.35 & GLADE+ \\
                  $0.051-0.089$ & 0.98 & 0.33  & 0.75 &  0.46 & NED-LVS \\
                  \hline
                  \multicolumn{6}{c}{Simulated GRBs}\\
                  \hline
                  0.131 ($\beta=1$) & 0.97 & 0.35  & 0.35 &  0.35 & GLADE+ \\
                  0.071 ($\beta=2$) & 0.95 & 0.20  & 0.50 &  0.39 & GLADE+ \\
          \hline
        \end{tabular}
\end{table*}

\begin{figure*}
   \centering
   \plotone{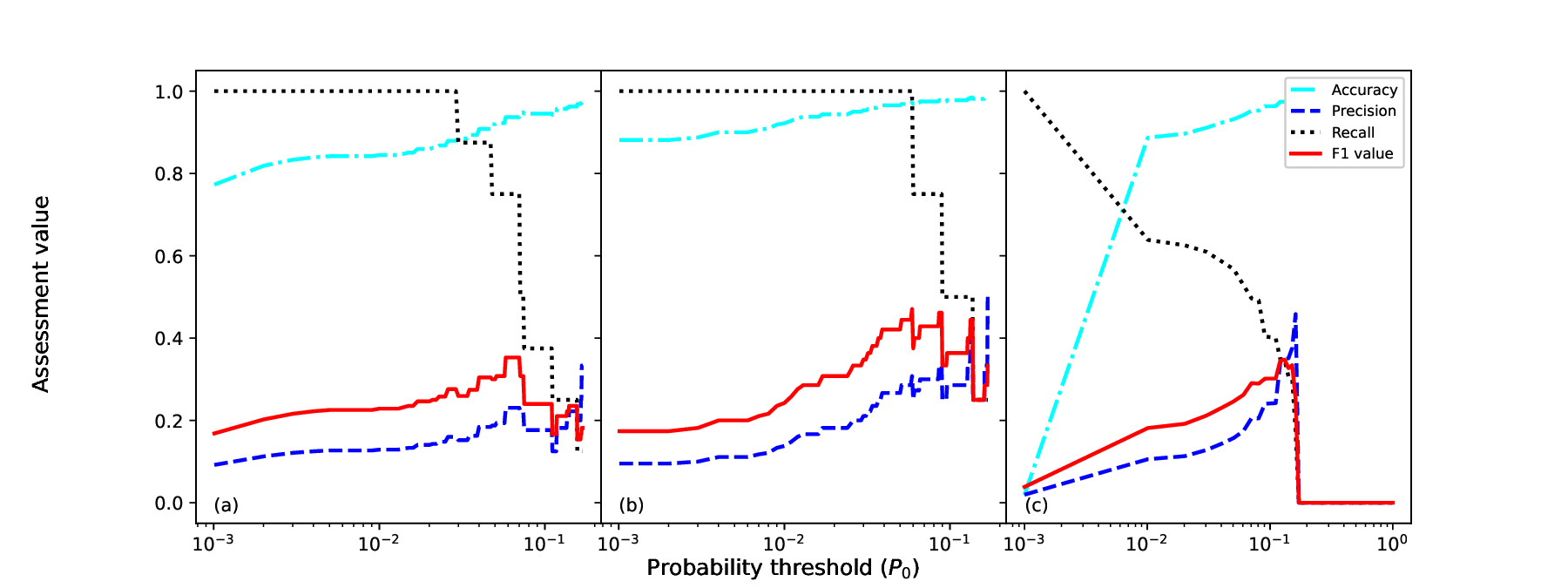}
   \caption{\it Panel (a): \rm
   Various assessments (see footnote 6 for the definitions) of the model plotted against the probability threshold $P_0$ when the observed GRB sample and the GLADE+ catalog are used. The optimal performance of the model can be determined by the peak of F1-score.
   \it Panel (b): \rm the same as panel (a) but for the case in which the NED-LVS catalog
   is used. \it Panel (c): the same as panel (a) but for the simulated GRB samples. 
}
\label{Fig4}
\end{figure*}

\subsubsection{Evaluation from Simulated GRB Samples}

The performance of the proposed model is additionally evaluated by 
a Monte-Carlo simulation with 1,000 random experiments. 
Each experiment is implemented by a random sampling of 100 local GRBs 
with $z<0.1$ and of 5~000 GRBs with $z>0.1$. The ratio of 100:5000 is roughly 
consistent with the statistics shown in the right panel of Figure 2.  

On the one hand, the simulated GRBs with 
$z>0.1$ are required to be uniformly distributed on the sky. On the other hand,     
the celestial positions of the simulated local GRBs are required to follow Eq. (1). 
Specifically, a sample of local
galaxies ($z<0.1$) following the galaxy mass distribution given in Eq. (2) are 
at first randomly extracted from a given galaxy catalog. 
Each extracted galaxy is then paired with an associated GRB, with the linear distance
between them is randomly sampled  according to the distribution given
in Eq. (3).

With the simulated samples, the model is then assessed by following the method used in 
Section 4.2.1. After an average of the 1,000 random experiments, panel (c) in Figure 4 
illustrates the dependence of the various
assessments on the probability threshold $P_0$, when the GLADE+ catalog is used. 
One can see from the figure the F1-score 
peaks at a large $P_0=0.131$, which means the model has the optimal performance with 
a small Recall$=0.35$. 

By definition, an assessment based on the F1-score means the 
Precision and Recall have equal importance. In practice, one, however, would like to 
reduce the fraction of missed local GRBs as much as possible, provided that the 
consumption of observational resources remains acceptable. This issue could be addressed by 
the generalized F-measurement\footnote{
\begin{equation}
  F_\beta = \frac{(\beta^2+1)\cdot\mathrm{Precision\cdot Recall}}{\mathrm{\beta^2\cdot Precision+Recall}}
\end{equation}
where $\beta$ is a parameter. When $\beta=1$, it reduces to the traditional F1-score. Precision 
(Recall) has larger weight in the case of $\beta>1$ ($\beta<1$).} $F_\beta$, peaking at different $P_0$ for different values of $\beta$.
After determining the peak of $F_\beta$,  
a dependence of the probability threshold $P_0$ on $\beta$ is shown in Figure 5.
Although $P_0$ decreases with $\beta$ generally,
two steps can be learned from the figure. One step occurs at $\beta\sim1$, and 
the other at $\beta\sim2$. Again, Table 2 lists 
the corresponding values of the assessments in both $\beta=1$ and $\beta=2$ cases. 
In the case of $\beta=2$, the optimal performance of the model has an increased Recall$=0.5$ and 
reduced Precision$=0.20$, which means a half of the true local GRBs would be missed by the 
model with a probability threshold $P_0=0.07$. 

\begin{figure}
   \centering
   \plotone{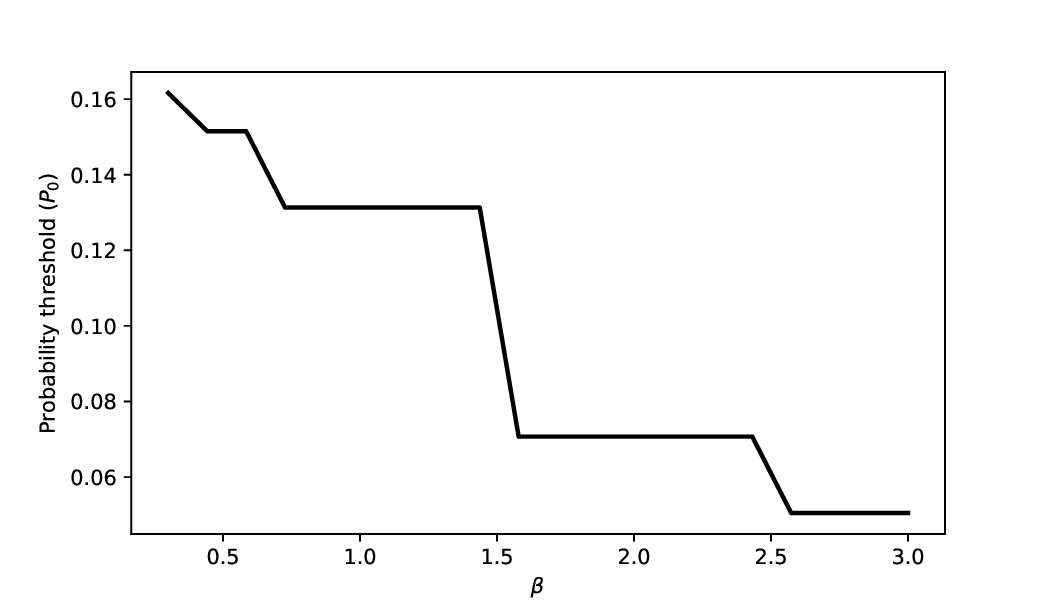}
   \caption{Probability threshold ($P_0$) determined by the peak of F-measurement 
   (see footnote 7) plotted 
   against $\beta$, when the model is assessed by the simulated GRB samples and the GLADE+ catalog.
}
\label{Fig5}
\end{figure}

\subsection{Comparison with Other Algorithms}

The results returned from the model proposed in this study are compared to two other 
algorithms recently reported in literature, by using the same observed GRB sample 
(i.e., the right panel in Figure 2).
On the one hand, by involving the
ALLWISE catalog (Cutri et al. 2013), 
Pan-STARRS catalog (Chambers et al. 2016), Hubble Source catalog (Whitmore et al. 2016)
and GLADE catalog (Dalya et al. 2018), only one association between GRB\,170817A and NGC\,4993 
has been claimed by the \texttt{Galclaim} tool\footnote{https://github.com/jgducoin/galclaim.} (Ducoin et al. 2023). The tool aims at identifying an association 
between a transient and corresponding host galaxy by the probability of chance alignment.

On the other hand, as an additional comparison, we run the tool of \texttt{astropath}\footnote{https://github.com/FRBs/astropath.} (Aggarwal et al. 2021) based on the GLADE+ catalog. The tool is designed to predict an association of a FRB and its host galaxy by posterior probability calculated based on the rule of Bayes. 
With the same observed GRB sample used in the current study, seven associations with 
a posterior probability $>0.9$ are in total predicted by the tool. 
There is, however, only one true association (i.e., GRB\,060505) 
among the seven events after a visual examination one by one.
The other 6 predicted associations are easily identified as being incorrect, 
because of the 
significant redshift difference between the observed afterglows and predicted host galaxies.

\section{Conclusion and Application}

By involving the masses and physical sizes of GRB's host galaxies, a model based on 
two-dimensional cross-match probability
is developed in this paper to identify local GRB candidates as soon as possible after 
the triggers and before spectroscopy.
The model is developed in the python 3.10 environment.
The packages of NumPy, SciPy (Virtanen et al. 2020), and sciki-learn
module (Pedregosa et al. 2011) are required for running the model.
With the GLADE+ and NED-LVS galaxy catalogs with a completeness of at least 
$\sim70\%$ up to $\sim300$Mpc,
the optimal performance of the model has a Precision of $0.23-0.33$ and
a Recall of 0.75 determined by the observed GRB sample. The optimal performance has
a reduced Recall of $0.35-0.50$ when the simulated GRB samples are adopted.

Figure 6 shows an example of implementation of the model in terms of 
a developed web service. In addition to the predicted association probability, 
the tool allows users to examine the predicted GRB-galaxy match visually,
and will be
integrated into the SVOM Science User Support System\footnote{https://www.svom.cn/suss/\#/home}
to help the SVOM burst advocators and GRB scientists to identify candidates of 
high-value source and to
reasonably allocate observational resources, especially spectroscopy, as soon as possible,
which is crucial for enhancing the scientific returns of the SVOM mission.
  
Due to the incorporation of galaxy mass (as well as its physical size) into the model,
the redshifts (either spectroscopic or photometric) or distances of galaxies are 
necessary in the model prediction. This is, however, a serious problem for the faint 
galaxies that either lack a redshift 
measurement in existing catalogs or are entirely excluded from the existing catalogs. This issue could be 
alleviated by estimating photometric redshifts (photo-$z$) by fitting their spectral energy distributions
if the multi-wavelength photometry is available. This approach is essential for a visual 
examination when a deep survey (e.g., DESI) is used. The photo-z estimation based on 
public code (e.g., \texttt{Hyperz},  Bolzonella et al. 2011) will be 
integrated into the web service in the next step.

\begin{figure}
   \centering
   \plotone{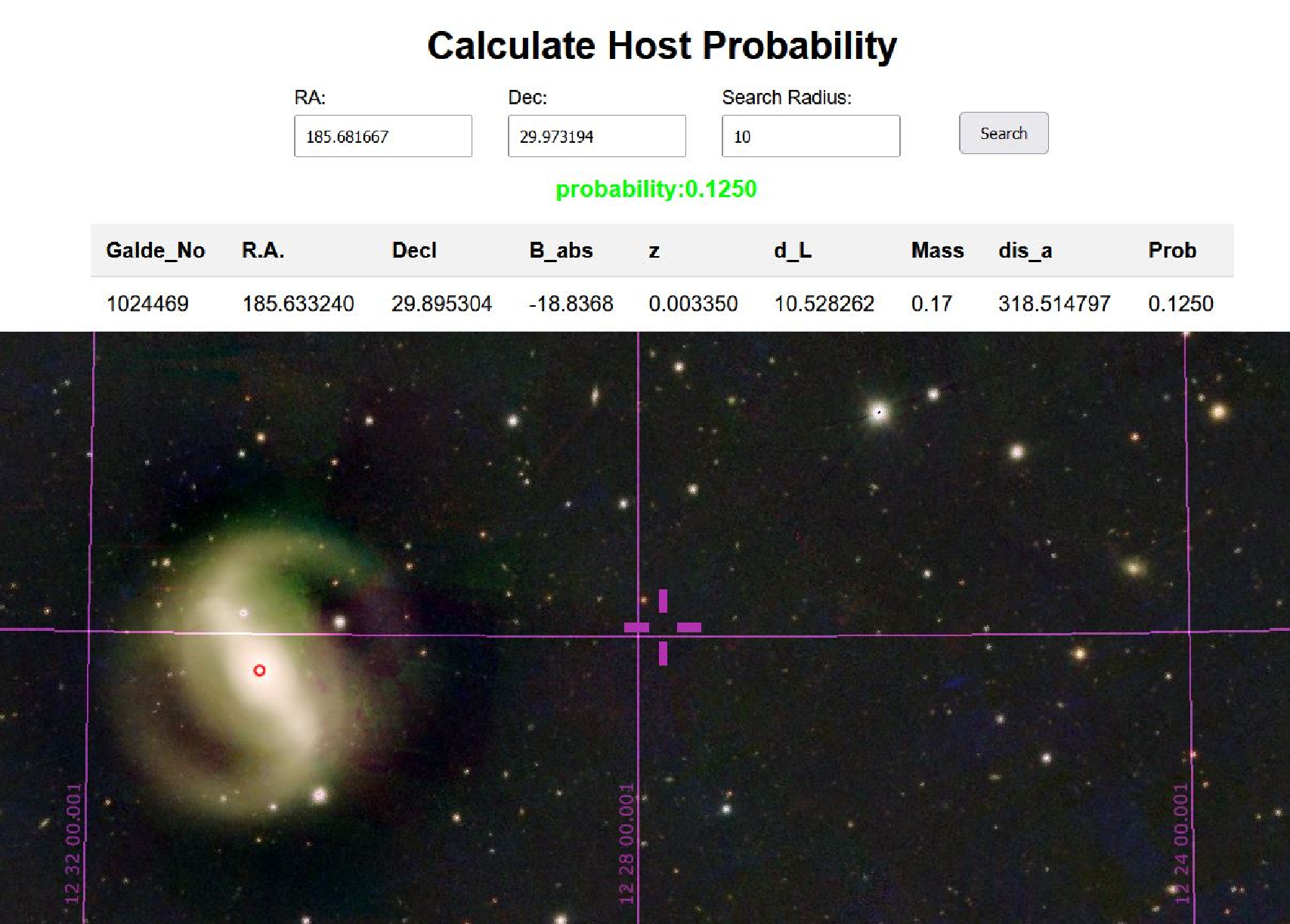}
   \caption{Implementation of the model of identifying nearby 
   GRBs by their most probable host galaxies. GRB~241218A with a \it Swift/\rm XRT enhanced 
   localization with an accuracy of arcsec (Beardmore et al. 2024) is used as an example. The calculated 
   possibility of 0.1250 shown in green means the value is larger than the probability
   threshold $P_0$ of our model. The positions of the 
   GRB and the galaxy with the highest cross-match probability are marked by the magenta cross and red circle, respectively. 
}
\label{Fig1}
\end{figure}

\acknowledgments
The authors are thankful for support from the National Key R\&D Program of China 
(grant No. 2024YFA1611700).
This study is supported by the Strategic
Pioneer Program on Space Science, Chinese Academy of Sciences,
grants XDB0550401. JW is supported by the National Natural Science Foundation of
China (Grants No. 12173009). 
EWL is supported by
the National Science Foundation of China (grant No. 12133003) and the Guangxi Talent 
Program (``Highland of Innovation Talents'').

\vspace{5mm}
\software{MATPLOTLIB (Hunter 2007), PYTHON/sklearn (Pedregosa et al. 2011)}
%


\end{document}